# Optical instability of the earth's atmosphere

Pavel G. Kovadlo, Olga S. Kochetkova
Institute of Solar-Terrestrial Physics, Russian Academy sciences, Siberian Branch, Irkutsk, Russia

## ABSTRACT

Meteorological data have been used to calculate refractive index fluctuations – the indicator of optical instability of the Earth's atmosphere. The calculations were made for standard pressure levels of the atmosphere in winter and summer. They are presented as distributions over the Earth's surface. The findings enabled us to determine preferred areas for astronomical observations as well as to compare astroclimate conditions of the world's largest observatories.

## 1. INTRODUCTION

This paper presents results of astroclimate investigations based on a great body of accumulated meteorological observations across the Earth. These data are not special and are therefore rather difficult to use (e.g., for estimating "seeing"). The major problem is the parameterization of optical turbulence processes. Preliminary studies of series of direct measurements of "seeing" and series of refractive index fluctuations ($\sigma N$) calculated from network radiosonde meteorological parameters demonstrated their high correlation over long periods with a correlation coefficient 0.6–0.95 [5]. An example is given in figure 1 presenting average long-term seasonal changes of "seeing" fluctuations (the angle of turbulence) photographed by the telescope AZT-7 (d = 20 cm) in Novosibirsk [2]. The thick line is the integral value calculated for the 20-km atmospheric layer. The thin lines indicate seasonal "seeing" dynamics in seconds of arc at various zenith distances.

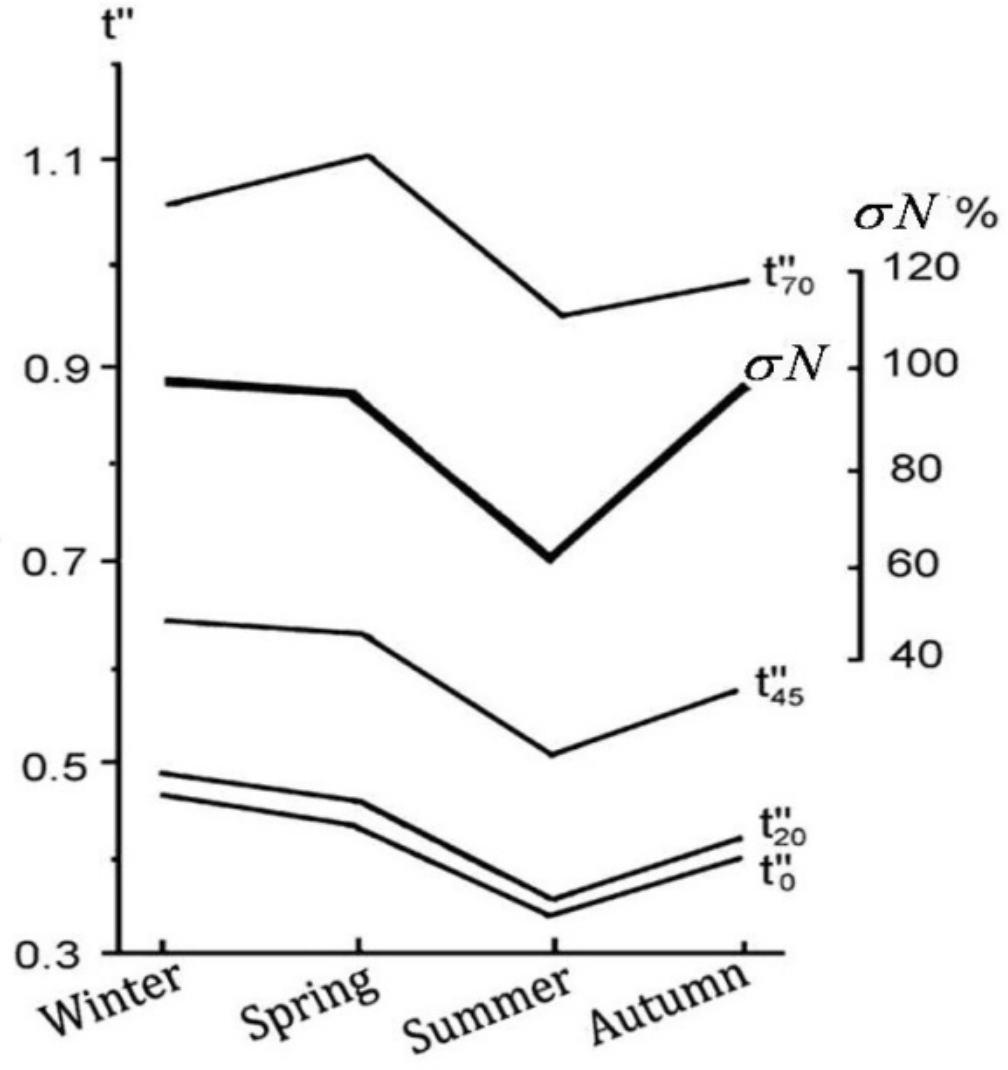

Figure 1: Seasonal changes of refractive index fluctuations

Such correlation in changes of large- and small-scale turbulence suggests definite successes in developing techniques of their mutual estimation, also in the spectral range of optical turbulence.

## 2. THE RESULTS AND DISCUSSION

The refractive index fluctuations are defined by the equation [3]:

$$\sigma N = \frac{A \cdot P}{T^2} \cdot \sigma T,$$

where A – empirical coefficient depending on the wavelength of light and for the green line is equal 80 (K/mb), P – mean monthly pressure (mb), T – mean monthly temperature (K), σT – the standard deviation of the temperature (K ).

The next figures show calculation results as distributions of σN at 700, 300 and 50 hPa from various meteorological archive data, including the NCEP/NCAR Reanalysis [1] ones. They have been compared with reference radiosonde measurements [4]. An example of the comparison results is given in figure 2. The isolines denote the percent of deviations calculated from NCEP/NCAR Reanalysis archive data and reference data at 850 hPa. The largest deviations for some stations were 21%.

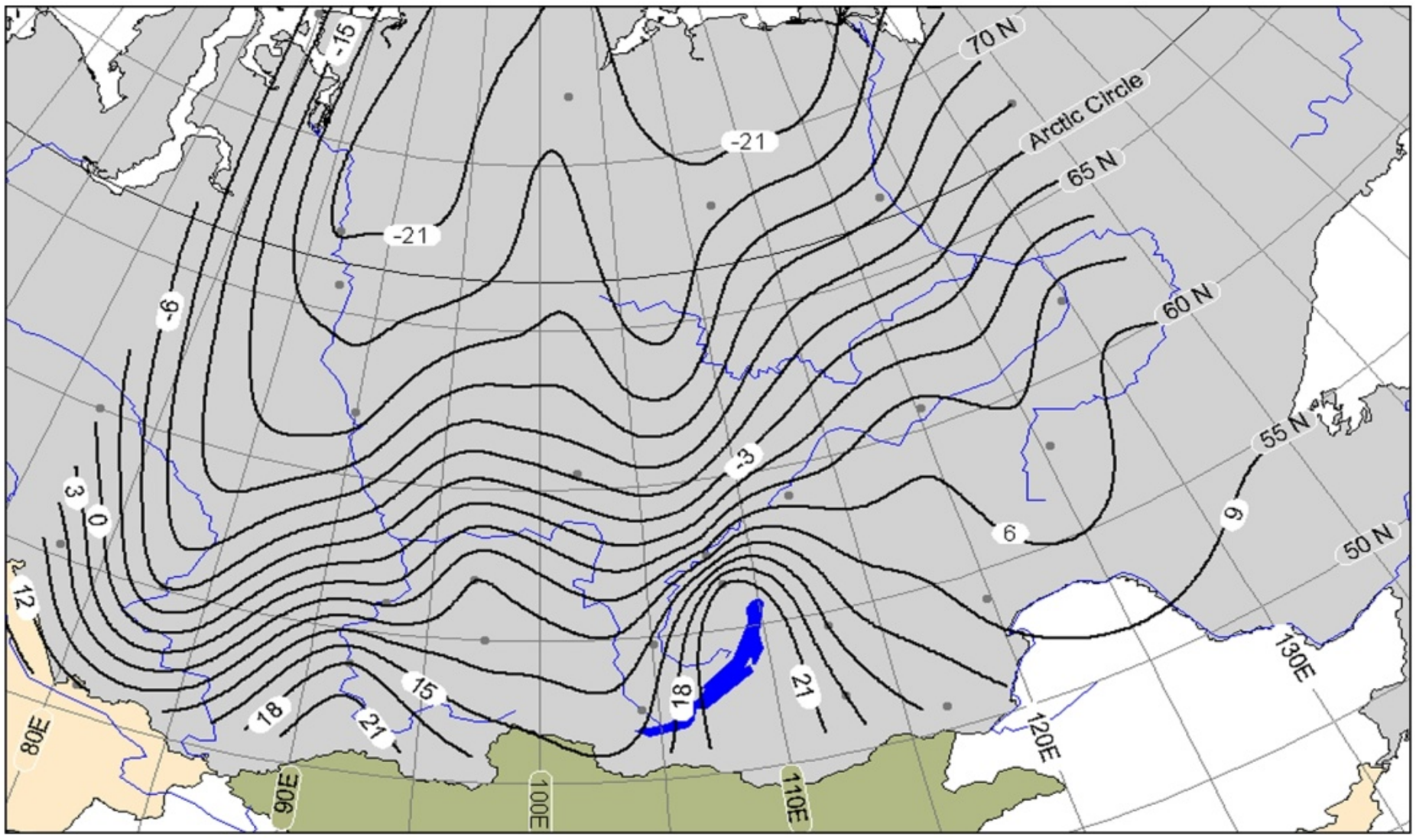

Figure 2: The relative deviation of σN, calculated from NCEP/NCAR Reanalysis data, from those obtained from reference data at 850 hPa

Figure 3 a) presents the calculation results as distributions of σN at 700 hPa (~ 3 km) in December–February, and figure 3 b), June–August over the period 1950–2009. These figures allow a quantitative comparison of the optical turbulence in different regions of the Earth.

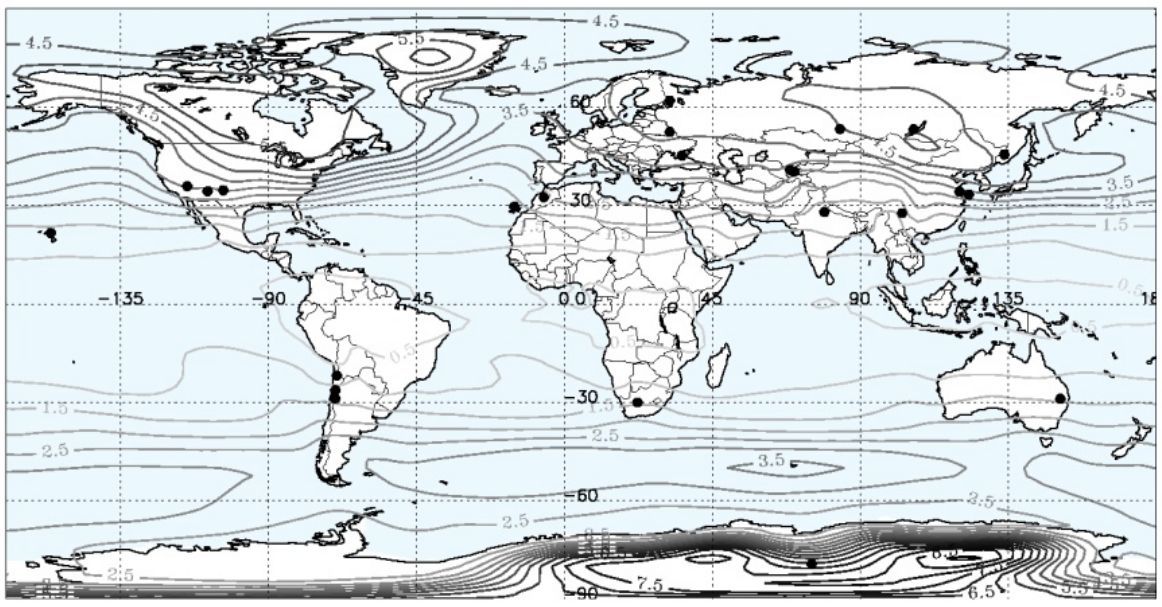

a)

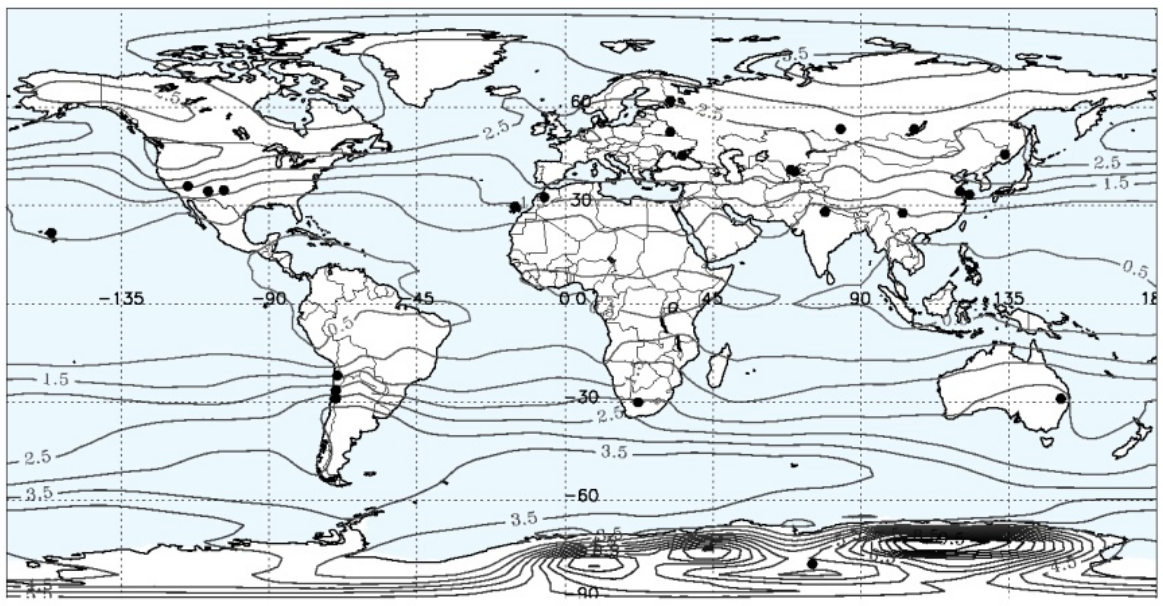

b)

Figure 3: The distribution of σN at 700 hPa a) December–February; b) June–August

The black points are large astronomical observatories and optical telescopes. The smallest values of σN are localized along the equator in the ± 30° zone. In June–August, this zone expands northward; in December–February, it contracts from the north. It is worth noting that it does not expand southward. Thus, the distribution depends on the season and latitude. The southern hemisphere in low latitudes has smaller values than the northern one there.

Figure 3 also shows that locations of observatories and large telescopes are characterized by small values of σN. It is rejoiceful that all astroclimate investigations were not in vain.

More detailed characteristics can be obtained from large-scale distribution patterns which can be constructed for a selected area. Figure 4 presents the pattern of distribution at 300 hGPa calculated from NCEP/NCAR Reanalysis archive data. Here it is possible to denote regions with large and small values.

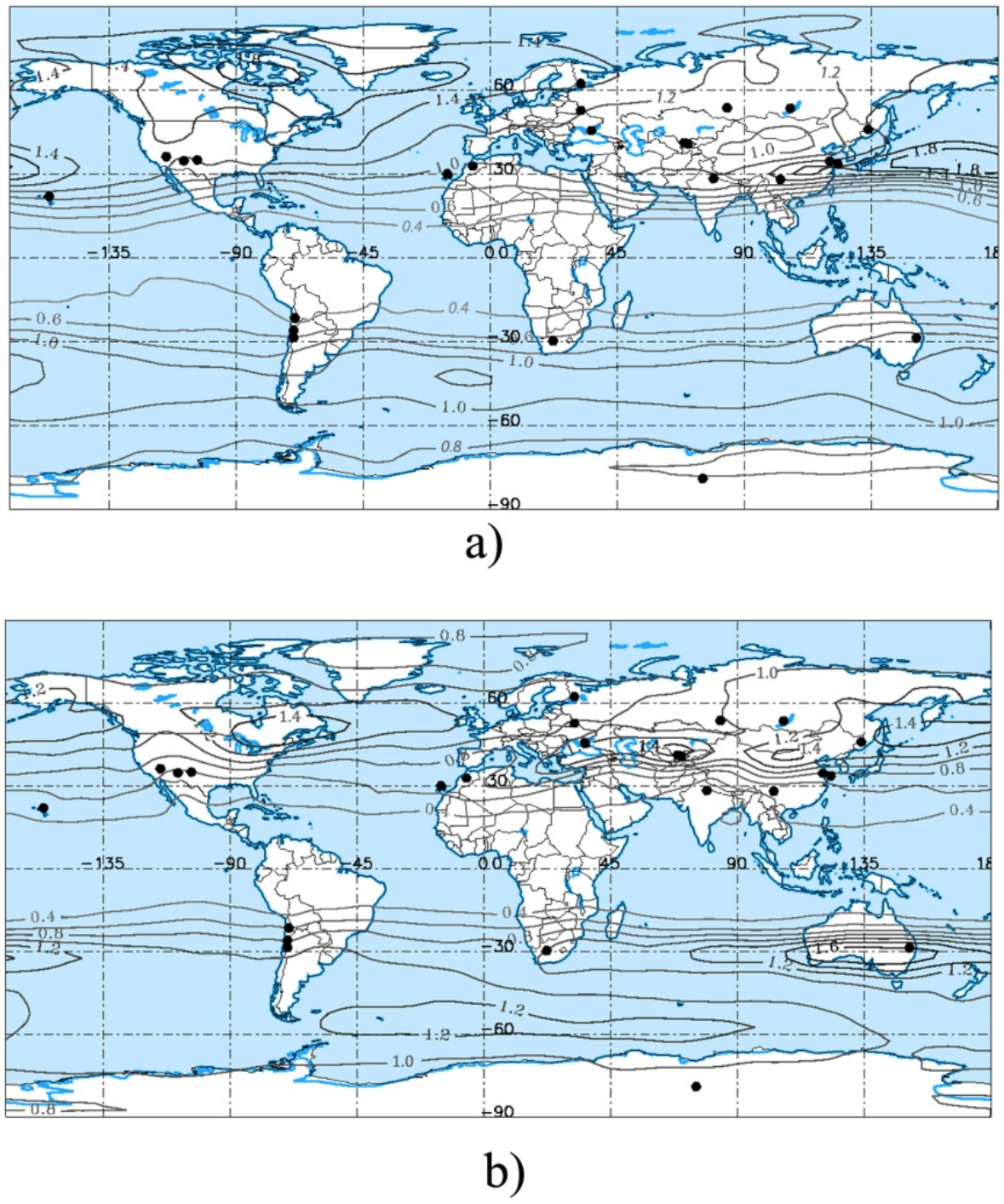

Figure 4: The distribution of σN at 300 hPa: a) December–February; b) June–August

This level (a height of 9 km) is interesting for the analysis of seeing characteristics first of all because it characterizes the under tropopause layer with the increased air turbulization being observed. The figures illustrate that the distribution as opposed to low levels (700 hPa) is less dependent on the underlying surface; the isolines are zonally directed for the most part. However, the most notable are the localized-in-space regions of maxima, in winter it is northern Canada and Greenland; in summer, southern Australia. Minimum values are recorded at the latitude from 150 N to 150 E in June–August.

To have an idea of the spatial distribution of optical turbulence in upper atmospheric layers, we constructed maps (Fig. 5) at 50 hPa (a height of 20 km; e.g., in the stratosphere).

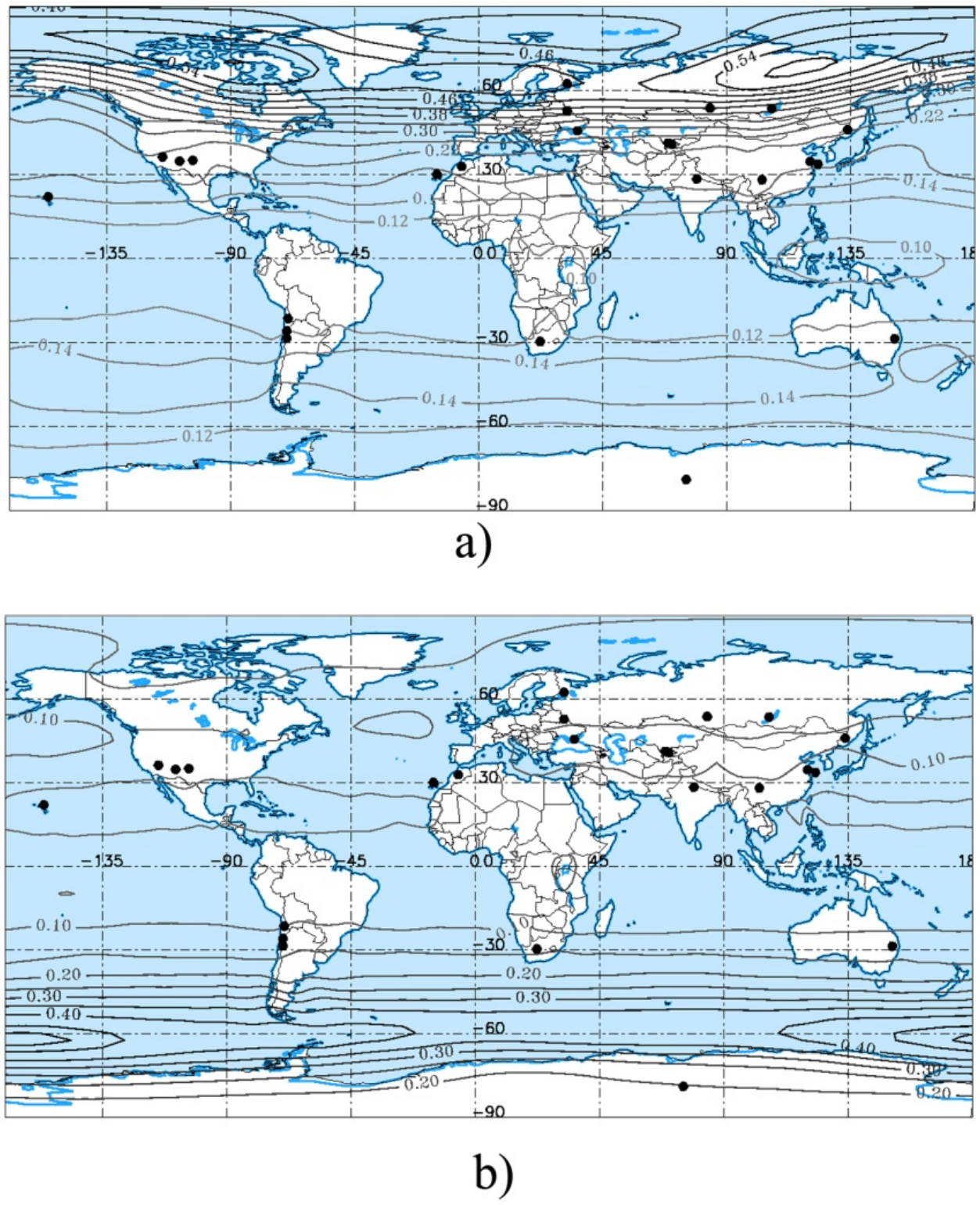

Figure 5: The distribution of σN at 50 hPa: a) December–February; b) June–August

A key feature of the distribution in this layer is the well pronounced latitude dependence. Maximum values are registered in the "winter" hemisphere; minimum ones, in the "summer" hemisphere. Such peculiarities are largely due to geophysical properties of the Earth; namely, the sphericity and the axial inclination to the plane of the ecliptic.

To examine astroclimate conditions of the world's largest observatories in more detail, we constructed a comparison table (Table 1) listing changes of refractive index fluctuations in height in January and July. Blue colour marks minimum among observatories at this level; red colour, maximum ones.

Table 1: Changes of refractive index fluctuations in height in January and July

| *Observatory* | Geographic coordinates; altitude above sea level (m) | January | | | | | July | | | | |
|---|---|---|---|---|---|---|---|---|---|---|---|
| | | 1000 mb (~ 0 км) | 700 mb (~ 3 км) | 300 mb (~ 9 км) | 50 mb (~ 20 км) | Mean | 1000 mb (~ 0 км) | 700 mb (~ 3 км) | 300 mb (~ 9 км) | 50 mb (~ 20 км) | Mean |
| 1. Mauna Kea (Hawaii) | 19° 49' N, 155° 28' W; **4205** | **0.78** | 1.22 | 0.93 | 0.13 | 0.77 | **0.40** | 0.91 | 0.46 | **0.05** | 0.46 |
| 2. CTIO Cerro Tololo Inter-American Obs. (Chili) | 29° 15′ S, 70° 44′ W; **2215** | 0.98 | 1.08 | 0.57 | 0.10 | 0.68 | 1.97 | 2.40 | 0.85 | 0.12 | 1.34 |
| 3. ESO (VLT) Cerro Paranal (Chili) | 24° 37′ S, 70° 24′ W; **2635** | 1.01 | **0.86** | **0.46** | 0.09 | **0.61** | 2.26 | 1.95 | 0.75 | 0.11 | 1.27 |
| 4. ESO La Silla Obs.(Chili) | 29° 05′ S, 70° 44′ W; **2400** | 2.16 | 1.75 | 0.97 | 0.20 | 1.27 | 0.60 | 0.57 | 0.53 | 0.06 | 0.44 |
| 5. La palma Obs. (Canary Islands) | 28°40′ N, 17°52′ W; **2400** | 2.64 | 3.58 | **1.63** | 0.51 | 2.09 | 1.00 | 2.16 | 1.14 | 0.08 | 1.10 |
| 6. NLST (Indian Astronomical Observatory) | 34°08′ N, 77°33′ E; **4517** | **9.06** | 4.04 | 1.19 | **0.52** | **3.70** | 3.70 | 2.35 | 0.99 | 0.07 | 1.78 |
| 7. Siding Spring Obs. (Australia) | 31°16′ S, 149° 03′ E; **1165** | 2.38 | 1.27 | 1.12 | 0.13 | 1.23 | 2.46 | 2.02 | **1.71** | 0.13 | 1.58 |
| 8. SALT, SAAO (Sutherland, South Africa) | 32°22′ S, 20°48′ E; **1783** | 2.19 | 1.49 | 0.73 | 0.12 | 1.13 | 2.53 | 2.29 | 1.01 | 0.15 | 1.50 |
| 9. RIDGE A ( Antarctica) | 81°5′ S, 73°5′ E; **4053** | 4.33 | 3.27 | 0.74 | **0.08** | 2.11 | **6.07** | **4.57** | 0.87 | **0.19** | **2.93** |
| 10. NSO Sacramento Peak (USA) | 32°22′ N, 116° 51′ W; **2800** | 4.14 | 2.99 | 1.14 | 0.21 | 2.12 | 1.86 | 1.23 | 0.51 | 0.07 | 0.92 |
| 11. NSO Kitt Peak (USA) | 32°18′ N, 110° 35′ W; **2095** | 5.94 | 2.59 | 1.17 | 0.19 | 2.47 | 1.17 | 0.71 | 0.59 | 0.06 | 0.63 |
| 12. Big Bear (California, USA) | 33° 07′ N, 116°51′ W; **2067** | 2.46 | 2.74 | 1.25 | 0.21 | 1.67 | 1.77 | 1.08 | 0.64 | 0.06 | 0.89 |
| 13. ISON-Ussuriysk Observatory (Russia) | 43°41 N, 132°09 E; **274** | 4.94 | 3.93 | 1.32 | 0.21 | 2.60 | 1.80 | 1.50 | 1.30 | 0.10 | 1.18 |
| 14. Large Vacuum Solar Telescope (Irkutsk, Russia) | 51°50′ N, 104°53′ E; **650** | 5.90 | **4.59** | 1.05 | 0.35 | 2.97 | 2.76 | 2.22 | 1.12 | 0.09 | 1.55 |
| 15. Altai Optical Laser Center | 51°21′ N, 82°09′ E; **385** | 6.45 | 3.86 | 1.07 | 0.36 | 2.94 | 2.82 | 2.16 | 0.98 | 0.10 | 1.52 |
| 16. Pulkovo Obs. (Russia) | 59°46′ N, 30°19′ E; **75** | 6.41 | 4.04 | 1.39 | 0.50 | 3.09 | 2.46 | 1.96 | 1.06 | 0.08 | 1.39 |
| 17. Main Astronomical Obs. (Ukraine) | 50°21′ N, 30°29′ E; **213** | 5.75 | 3.75 | 1.20 | 0.36 | 2.77 | 2.39 | 1.80 | 0.96 | 0.10 | 1.31 |
| 18. Crimean Astrophysical Observatory (Ukraine) | 44° 43′ N, 34° 00′ E; **600** | 3.33 | 3.28 | 1.08 | 0.26 | 1.99 | 1.90 | 1.85 | 1.31 | 0.12 | 1.30 |
| 19. Majdanak Obs. (Uzbekistan) | 38° 67′ N, 66° 89′ E; **2593** | 3.92 | 2.68 | 1.04 | 0.22 | 1.97 | 1.92 | 1.64 | 1.05 | 0.12 | 1.18 |
| 20. Sanglok Obs. (Tadjikistan) | 38°15′ N, 69°13′ E; **2300** | 3.92 | 2.68 | 1.04 | 0.22 | 1.97 | 1.92 | 1.64 | 1.05 | 0.12 | 1.18 |
| 21.Yunnan Astronomical Obs. (Yunnan, China.) | 25° 44′ N, 102° 01′ E; **2014** | 1.34 | 1.14 | 1.10 | 0.21 | 0.95 | 0.54 | **0.48** | **0.37** | 0.07 | **0.37** |
| 22. Purple Mountain Obs. (Nanjing, China.) | 32°03′ N, 118°46′ E; **267** | 4.09 | 2.70 | 1.61 | 0.19 | 2.15 | 1.29 | 0.95 | 0.65 | 0.09 | 0.75 |
| 23. Shanghai Astronomical Obs. (Shanghai, China) | 31°13′ N, 121°28′ E; **100** | 3.63 | 2.36 | 1.52 | 0.17 | 1.92 | 0.93 | 0.74 | 0.58 | 0.07 | 0.58 |

One can see that varies depending on the level, and the selected observatories do not have extreme values at all levels. However, in spite of this, there are observatories with largest and smallest among them. The former in January are NLST (Indian Astronomical Observatory), La palma and at 700 hPa the maximum σN was recorded by the Large Vacuum Solar Telescope; in July – RIDGE A, Siding Spring, NLST (Indian Astronomical Observatory) at 200 hPa. Minimum σN can be observed in January at ESO (VLT) Cerro Paranal, RIDGE A and at 1000 hPa at Mauna Kea; in July – at Yunnan Astronomical Obs., at low and upper level at Mauna Kea. As already noted, the distribution is inhomogeneous in height.

To understand the vertical structure σN, we constructed vertical profiles of these characteristics.

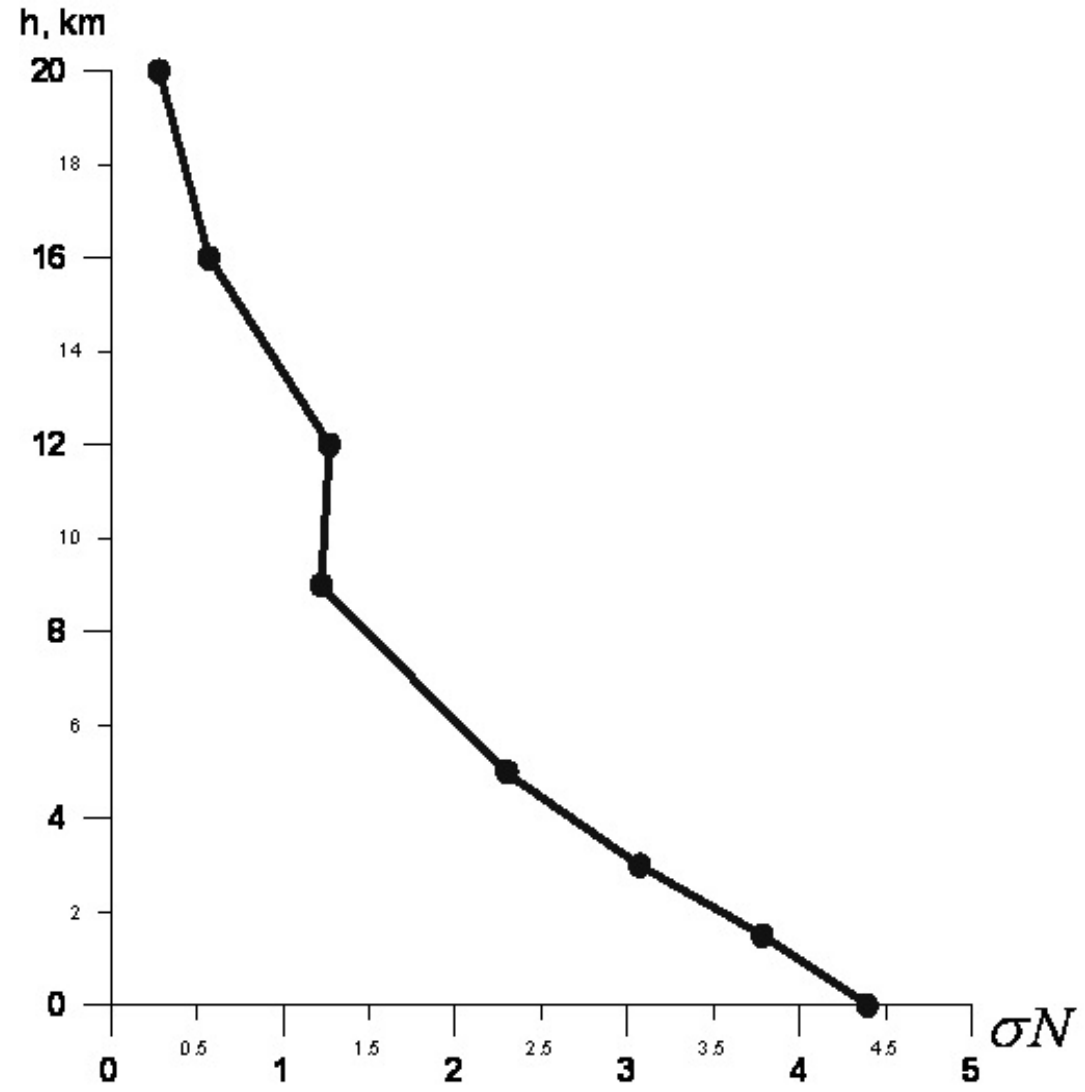

Figure 6: Vertical profiles σN

The values were averaged within the northern and southern hemispheres for January and July. By and large, refractive index fluctuations decay with increasing height that depends on the distance from the underlying surface, their source. But in the layer between 9 and 12 km, increases slightly. This is associated with the increased turbulent energy of the air layer under the tropopause.

## 3. CONCLUSION

The application of the proposed technique of numerical zoning enables us to analyze the large-scale pattern of optical instability across the Earth as well as to determine new sites with the lowest values of refractive index fluctuations. Besides, the technique offers complementary possibilities of planning a great number of events far beyond astronomer's field of interest.

## REFERENCES

1. E. Kalnay et al., “The NCEP/NCAR 40-year reanalysis project”, Bull. Amer. Meteor. Soc., Vol.77, pp. 437-471, 1996.
2. E.M. Afanasieva, The quality of images of stars from observations of 1961 – 1963 years in Novosibirsk, Novosibirsk: Nauka, 1967.
3. V.I. Tatarsky, Wave propagation in turbulent atmosphere, Moscow: Nauka, 1967
4. I.G. Guterman, “New aeroclimatic reference free atmosphere over the USSR. The temperature”, Vol II, Moscow: Gidrometeoizdat, P.150, 1980.
5. P. G. Kovadlo “Some characteristics of the yearly mean variation of scattered solar radiation during clear sky conditions over the CIS territory and optical instability of atmosphere”, Geophysics & Astronomy, Vol.3, pp.75-79, 2007